\documentclass[a4paper,12]{article}
\usepackage[utf8]{inputenc}
\usepackage{pifont,textcomp}
\usepackage{amsfonts}
\usepackage{amsmath}

\usepackage{color}
\usepackage{graphics}

\usepackage{amscd}
\usepackage{array}
\usepackage{multirow}
\usepackage{mathrsfs}
\usepackage{amssymb}
\usepackage{amsthm}
\newcommand{\beqn}{\begin{eqnarray}}
\newcommand{\eeqn}{\end{eqnarray}}
\newcommand{\beq}{\begin{equation}}
\newcommand{\eeq}{\end{equation}}
\newcommand{\bpro}{\begin{proposition}}
\newcommand{\epro}{\end{proposition}}
\newcommand{\blem}{\begin{lemma}}
\newcommand{\elem}{\end{lemma}}
\newcommand{\bdfn}{\begin{definition}}
\newcommand{\edfn}{\end{definition}}
\newcommand{\bcor}{\begin{corollary}}
\newcommand{\ecor}{\end{corollary}}
\newcommand{\bthm}{\begin{theorem}}
\newcommand{\ethm}{\end{theorem}}
\newcommand{\bex}{\begin{example}}
\newcommand{\eex}{\end{example}}
\newcommand{\brmq}{\begin{remark}}
\newcommand{\ermq}{\end{remark}}
\newcommand{\benum}{\begin{enumerate}}
\newcommand{\eenum}{\end{enumerate}}
\newcommand{\bitem}{\begin{itemize}}
\newcommand{\eitem}{\end{itemize}}

\title{  Photon-added   Barut-Girardello like coherent states of time-dependent Landau  problem \\{\it \large {Dedicated to the memory of Lev Davisdovich Landau}} }

 \author{Latévi M. Lawson$^{a,b}$, Komi Sodoga$^b$ and Gabriel Y. H. Avossevou$^c$  \\
 \space\\
 \textit{ }
  \textit{${}^{a}\!$African Institute for Mathematical Sciences (AIMS) Ghana,}\\\textit{Summerhill Estates, East Legon Hills,}\\\textit{P.O. Box LG DTD 20046, Legon, Accra, Ghana}\\\\
  \textit{${}^{b}\!$Laboratoire de Physique des Matériaux et des Composants} \\\textit{à Semi-Conducteurs,
  	02 BP 1515 Lomé, Togo}\\\\
 \textit{${}^{c}\!$Institut de Mathématiques et de Sciences Physiques (IMSP)},\\
\textit{ Laboratoire de Recherche en Physique Théorique (LRPT)}\\
 \textit{01 BP 613 Porto-Novo, Rep. du Bénin}\\\\
 latevi@aims.edu.gh$^{a}$; ksodoga@univ-lome.tg$^b$ and\\ gabriel.avossevou@imsp-uac.org$^c$}

\begin{document}

\maketitle

\begin{abstract}
Recently, we have   determined  the spectrum and the wave functions of the Hamiltonian 
 of a Landau particle with time-dependent mass and frequency undergoing the influence of a uniform time-dependent electric field  [J. Math. Phys. 56, 072104 (2018)]. In the  present  paper  we extend the study of this model that we name the time-dependent Landau  problem into the context of coherent states. By  means  of the traditional  factorization method of the  eigenfunctions of this system expressed in terms of the generalized Laguerre polynomials, we derive the generators of the $su(1,1)$ Lie algebra and we construct the coherent states   à la Barut-Girardello. These states 
 are shown to satisfy the Klauder's mathematical requirement to build coherent states and 
 some of their  statistical properties are  calculated and analyzed. We find that these states are sub-Poissonian in nature. We show that, addition of photons from these coherent states, increases the    statistical properties and changes the mathematical properties of these states.
  
 \end{abstract}

\section{Introduction}
The study of coherent states has remained over the past four decades a constant source of fascination,  inspiration and innovation in 
different branches of physics. In particular, they have found considerable applications in quantum optics \cite{1',2} and quantum information
\cite{2,4}. They were first discovered in connection with the quantum harmonic
oscillator by Schr\"{o}dinger in 1926, who referred to them as states of minimum uncertainty product \cite{5}. 
The importance of coherent states was put forward by Glauber  in the framework of quantum optics in early 1960's \cite{6,7}.
According to Glauber, these states
can globally be constructed using any of the following three procedures :  (i) annihilation
operator eigenstates (ii) displacement operator technique (iii) minimum uncertainty states. However, these three approaches
are generally nonequivalent and only in the case of standard harmonic oscillator coherent
states obtained using any of the three approaches are equivalent. The same states were also reintroduced by
Klauder  who investigated their mathematical properties \cite{8,9}. He has noted that, these states must satisfy the following minimum conditions:
(iv) normalizability, (v) continuity in the label and (vi) existence of a resolution of unity with a positive definite weight function.
Nine years later, these coherent states introduced by Glauber have inspired respectively Barut-Girardello \cite{10} and Perelomov \cite{11} in constructing the coherent states for $su(1,1)$ Lie algebraic group basing on the procedures (i) and (ii) respectively.
These states also satisfy the Klauder's minimum conditions (iv,v,vi) and have interesting applications in quantum optics, quantum computation
and quantum mechanics \cite{12,13,14,15,16}.

On the other hand, the addition of photons from  the coherent states previously introduced by Glauber were objects of 
extensive studies both in experimental \cite{17,18,19} and theoretical \cite{20,21,22} frameworks.
These states originally introduced by   Agarwal and Tara \cite{23,24} have received much attention from quantum optics and  information \cite{25,26,27} for 
their statistical properties such as the photon-number distribution, the Mandel parameter, the squeezing parameter, the Wigner function etc. Various 
generalizations of these states taking into account their  statistical properties  were also performed \cite{28,29,30,31}. Recently, one of us addressed 
 conjonctly the   study  of photon added states into the generalized associated hypergeometric coherent states
  \cite{32} and   into a full characterization of  shape invariant potentials using algebraic approach based on the supersymmetric
 quantum mechanics \cite{33}.  The original construction of photon added coherent states (PACSs)  \cite{23,24}
 obtained  from the  conventionnal coherent states based on the Weyl-Heisenberg group
has been extended to a number of Lie groups with square integrable representations. In this sens, Popov constructed   and analytically discussed the statistical properties  of the photon-added   Barut-Giraldello  coherent states (PABGCSs) for the pseudoharmonic oscillator \cite{34}. Daoud extended this construction for exactly solvable
 Hamiltonian on Gazeau-Klauder and Klauder-Perelomov coherent states \cite{35}. Recently, the mathematical and statistical properties  of  PACSs  are  studied  for  the 
 $SU(2)$ coherent states \cite{21,36}.  
    We extend  these results  into   our  recently model studied in \cite{1} namely the   time-dependent Landau  problem (TDLP). To do so, we organise the paper as follows.
  
  In section \ref{section2}, we outline the fundamental  aspects of the quantization of the TDLP based on the Lewis-Riesenfeld procedure \cite{13,37}. 
  We recall  the spectra and the  eigenfunctions of the  different operators that describe the  system. In
  section \ref{section3}, we  derive  from the  solution of  the system the hidden  generators of the $su(1,1)$ Lie  algebra.
 We proceed by  the factorization method as developped in \cite{13,38,39}  to find  the hidden symmetry of the system and  derive 
 from the eigenfunctions the related raising and lowering operators which generate the $su(1,1)$ Lie algebra.
 Section \ref{section4} is devoted to the  construction of  the $SU(1,1)$ coherent  
 states   $\grave{\textrm{a}}$ la Barut-Girardello,  the   study of  their mathematical properties
 (the non-orthogonality, the continuity, the identity resolution) and to their
 statistical properties (the photon mean number, the photon distribution, the intensity correlation, the  Mandel parameter and
 the Wigner distribution functions). The  section \ref{section5}  presents in detail the  construction of photon added Barut-Giraldello like coherent states
 (PABGLCSs). We then check the effect of adding  
 photons on   the above mathematical and statistical properties.
  Finally, we  conclude the paper in section \ref{section6}.

\section{Review of  the model} \label{section2}
The coming january 22, 2021, the world scientific  will celebrate the 113th birth  anniversary of Soviet physicist Lev Landau. During   the 20th century, Lev Landau made   some of the  most  significant  discoveries in physics. In  quantum mechanics, he  is  known for the problem of  quantization of the cyclotron orbits of  charged particles in constant magnetic field. As a result, the charged particles can only occupy orbits with discrete energy values, called Landau levels. The Hamiltonian of this system is 

\begin{eqnarray} \label{eq2.1}
H&=&\frac{1}{2M}\left[\vec p-q\vec A(\vec x)\right]^2.
\end{eqnarray}
where $ \vec p$ is the canonical momentum operator, $\vec A$ is the electromagnetic vector potential,  which is related to the magnetic field  $\vec B$ by  $A_i(x_i)= -\frac{1}{2} B \epsilon_{ij} x_j$ (with $i, j = 1, 2$) and $M$ is the constant mass of the particle.
Recently,  we have extended the Hamiltonian of this system (1) to the case of 
time-dependent mass $M(t)$ and harmonic frequency $\omega(t)$ under the influence of a uniform time-dependent electric field $E(t)$ \cite{1}. This system is  described by the Hamiltonian
\begin{eqnarray} \label{eq2.1}
 H(t)&=&\frac{1}{2M(t)}\left[\vec p-q\vec A(\vec x)\right]^2+\frac{1}{2}M(t)  \omega^2(t)\vec x^2 +q\varphi(t).
\end{eqnarray}
In the symmetric gauge, the vector potential and the time-dependent
scalar potential are given by  $A_i(x_i)= B \epsilon_{ij} x_j$ and $\varphi(x_i ,t)=E_i (t)x_i$, respectively (with $i, j = 1, 2$).
The Hamiltonian of the system is rewritten as follows
\begin{eqnarray} \label{eq2.2}
 H(t) &=& \frac{1}{2M(t)}\left[p_1+ \frac{q}{2}  Bx_2\right]^2+\frac{1}{2M(t)}\left[p_2-\frac{q}{2}  Bx_1\right]^2\cr
    && +\frac{1}{2} M(t) \omega^2(t)(x_1^2+x_2^2) -q[E_1(t)x_1+E_2(t)x_2].
\end{eqnarray}
Considering the following changes of variables 
\begin{eqnarray} \label{eq2.3}
 x&=&x_1+\frac{qE_1(t)}{M(t)\omega^2(t)},\,\,\, y=x_2+\frac{qE_2(t)}{M(t)\omega^2(t)},\\
 p_x&=&p_1-\frac{q^2BE_2(t)}{2M(t)\omega^2(t)},\,\,\, p_y=p_2-\frac{q^2BE_1(t)}{2M(t)\omega^2(t)},
\end{eqnarray}
and get the transformed  Hamiltonian  in its simple  form rewritten as follows
\begin{equation} \label{eq2.4}
 H(t)=\frac{1}{2M(t)}(p_x^2+p_y^2)+\frac{\Omega^2(t)M(t)}{2}(x^2+y^2)-\frac{\omega_c (t)}{2}L_z-\frac{q^2E^2(t)}{2M(t)\omega(t)},
\end{equation}
where $L_z=xp_y-yp_x$ is the angular  momentum, $\omega_c(t)=\frac{qB}{M(t)}$ is the cyclotronic frequency of oscillation,
$\Omega(t)=\sqrt{\omega^2(t)+\frac{1}{4}\omega_c^2(t)}$ is the general frequency of oscillation,
$ E(t)=\sqrt{E_1^2(t)+E_2^2(t)}$ and $\hbar=1$.

 Since the Hamiltonian involves time-dependent parameters, we used the so-called Lewis and Riesenfeld method 
\cite{13} to construct
a Hermitian operator $I(t)$  as follows \cite{1}
\begin{equation} \label{eq2.5}
 I(t)=\frac{1}{2}\left[\frac{\kappa^2}{\rho^2}x^2+\frac{\kappa^2}{\rho^2}y^2+\left(\rho p_x-M\dot{\rho}x\right)^2+
 \left(\rho p_y-M\dot{\rho}y\right)^2\right],
\end{equation}
where $\kappa$ is a constant and the function $\rho$ is the solution of the so-called nonlinear  
modified Ermakov-Pinney equation \cite{40} which is
\begin{equation} \label{eq2.6}
 \ddot{\rho}+ \frac{\dot{M}}{M}\dot{\rho}+\Omega^2(t)\rho=\frac{\kappa^2}{M^2\rho^3}.
\end{equation}
The  eigensystems  of the invariant  are given as follows
\begin{eqnarray}\label{eq2.7} %\label{f2}
 \langle \phi_{n}^{|\ell|}(t)|I(t)|\phi_{n}^{|\ell|}(t)\rangle&=&\kappa(2n+|\ell|+1),\\
 \phi_{n}^{|\ell|}(r,\theta,t)&=& (-)^n\frac{(\kappa)^{\frac{1+|\ell|}{2}}}{\rho^{1+|\ell|}\sqrt{\pi}}\sqrt{\frac{n!}{\Gamma(n+|\ell|+1)}} r^{|\ell|}\times\cr&& e^{\left(iM
\frac{\dot{\rho}}{\rho}-\frac{\kappa}{\rho^2}\right)\frac{r^2}{2}}L_n^{|\ell|}\left(\frac{\kappa}{\rho^2}r^2\right)e^{i\ell\theta},
\end{eqnarray}
where $\ell=n_+-n_-$,\,\,\,$n=\min(n_+,n_-)=\frac{1}{2}(n_++n_--|\ell|)$, $\Gamma(u)$  Euler’s gamma function and $L_n^{|\ell|}\left(u\right)$
are the generalised Laguerre polynomials, while,
\begin{eqnarray} \label{eq2.8}
 r=\sqrt{x^2+y^2},\,\,\,\,\,\,\, e^{i\theta}=\frac{x+iy}{\sqrt{x^2+y^2}}.
\end{eqnarray}
The solution of the Schrödinger equation is obtained by taking the product
of the eigenfunction of the invariant operator and the exponential of the complex of the phase
function such as
\begin{eqnarray}\label{eq2.9}
\psi_n^{|\ell|}(r,\theta, t)&=&(-)^n\frac{(\kappa)^{\frac{1+|\ell|}{2}}}{\rho^{1+|\ell|}\sqrt{\pi}}\sqrt{\frac{n!}{\Gamma(n+|\ell|+1)}}
r^{|\ell|}\times\cr&& e^{\left(iM
\frac{\dot{\rho}}{\rho}-\frac{\kappa}{\rho^2}\right)\frac{r^2}{2}}L_n^{|\ell|}\left(\frac{\kappa}{\rho^2}r^2\right)e^{i\ell\theta} 
e^{i\gamma_n^\ell (t)}, 
\end{eqnarray}
where the phase  factor is given by
\begin{eqnarray}\label{eq2.10}
 \gamma_n^\ell (t)&=&-\frac{\kappa}{2}\left(2n+|\ell|+1\right)\int_0^t\frac{dt'}{M(t')\rho^2(t')}
 \cr&&-\frac{|\ell|}{2}\int_0^tdt' \omega_c(t')+\frac{q^2}{2}
 \int_0^t \frac{E^2(t')}{M(t')\omega(t')}dt'.
\end{eqnarray}
The Hamiltonian's expectation values are    given by the expression
\begin{footnotesize}
\begin{eqnarray*} %\label{eq2.10}
 \langle \phi_n^{|\ell|}(t)|H(t)|\phi_n^{|\ell|}(t)\rangle&=&\frac{1}{2\kappa}\left(M\dot{\rho}^2+\frac{\kappa^2}{M\rho^2}
 +M\Omega^2\rho^2\right)\left(2n+|\ell|+1\right)-\frac{|\ell|}{2}\omega_c
% \cr&&
 -\frac{q^2E^2}{2M\omega}.
 \end{eqnarray*}
 \end{footnotesize}
Since the eigenfunction of the system (\ref{eq2.8}) is expressed in terms of the generalized Laguerre functions, it is possible
to derive the hidden generators of the $su(1,1)$ Lie   algebra  through  factorisation of this  eigenfunction. In  what  follows, 
we firstly  construct  the raising and lowering operators from
the Hamiltonian’s eigenfunction which generates the $SU(1,1)$ hidden Lie  group. Secondly, we construct the 
 the Barut-Girardello Coherent states relative to this system. 
 Finally,  we introduce the associated PACSs  and study their properties.

\section{su(1,1)  Lie  algebraic treatment}\label{section3}
Before  constructing the  SU(1,1) coherent states to the system, it  is  important to review some usefull properties to 
the associated  Laguerre polynomials  in order to derive from the wavefunctions (\ref{eq2.7}) and (\ref{eq2.9}) the $su(1,1)$ Lie 
 algebraic treatment of the system. \\

\noindent The generalized Laguerre polynomials $L_n^\ell(u)$  with $\ell>0$ are defined  as \cite{41}
\begin{eqnarray}\label{eq3.1} % \label{la}
 L_n^{\ell}(u)&=&\frac{1}{n!}e^u u^{-\ell}\frac{d^n}{du^n}(e^{-u}u^{n+\ell})=\sum_{k=0}^n (-)^k \binom{n+\ell}{n-k}\frac{u^k}{k!}.
  \end{eqnarray}
  For $\ell=0,\,\,\,L_n^0(u)=L_n(u)$  and \,\,\,for $n=0,\,\,\,L_0^{\ell}(u)=1$. The generating function corresponding to the associated
  Laguerre polynomials is
  \begin{eqnarray} \label{eq3.2} % \label{j}
 J_{\ell}\left(2\sqrt{uz}\right)e^z(uz)^{-\frac{\ell}{2}}&=&\sum_{n=0}^\infty\frac{z^n}{\Gamma(n+\ell+1)}L_n^{\ell}(u),
\end{eqnarray}
where the $ J_\kappa(x)$ is the ordinary Bessel function of $\kappa$-order.\\

\noindent The orthogonality relation is 
\begin{eqnarray} \label{eq3.3}
 \int_0^{\infty}du e^{-u}u^{\ell}L_n^{\ell}(u)L_m^{\ell}(u)= \frac{\Gamma(\ell+n+1)}{n!} \delta_{nm}.                                                     
\end{eqnarray}
 The generalized Laguerre polynomials satisfy the following differential equation
 \begin{equation}
  \left[u\frac{d^2}{du^2}+(\ell-u+1)\frac{d}{du}+n\right]L_n^{\ell}(u)=0,
 \end{equation}
and the recurrence relations
\begin{eqnarray}  \label{eq3.4} %\label{r1}
 (n+1)L_{n+1}^{\ell}(u)-\left(2n+\ell+1-u\right)L_n^{\ell}(u)+\left(n+\ell\right)L_{n-1}^{\ell}(u)&=&0,\\ \label{eq3.41} % \label{r2}
 u\frac{d}{du}L_n^{\ell}(u)-nL_n^{\ell}(u)+(n+\ell)L_{n-1}^{\ell}(u)&=&0.
\end{eqnarray}
With respect to the  above equations, we rewrite the  eigenfunction of the invariant operator in equation (\ref{eq2.7}) in the form
\begin{equation}  \label{eq3.5}
 \phi_n^{\ell}(u,t)=N_n(\rho,\theta)\sqrt{\frac{n!}{\Gamma(n+\ell+1)}}u^{\frac{\ell}{2}}e^{-\frac{\varpi}{2}u}L_n^{\ell}(u),
\end{equation}
where $u=\frac{\nu}{\rho^2}r^2$, $N_n(\rho,\theta)=(-)^n\sqrt{\frac{\nu}{\pi\rho^2}}e^{i\ell\theta}$,\,\,\,\,
$\varpi=1-iM(t)\frac{\rho\dot{\rho}}{\nu}$ and $\Gamma(n)=(n-1)!$.\\\\
Based on the recurrence relations (\ref{eq3.4}) and (\ref{eq3.41}), we obtain the following equations
\begin{eqnarray}  \label{eq3.6}
 \left(-u\frac{d}{du}+\frac{\ell}{2}+n-\frac{\varpi}{2}u\right)\phi_n^{\ell}(u,t)&=&\sqrt{n(n+\ell)}\phi_{n-1}^{\ell}(u,t),\\ \label{eq3.7}
 \left(u\frac{d}{du}+\frac{\ell}{2}+n-\frac{\tilde \varpi}{2}u+1\right)\phi_n^{\ell}(u,t)&=&\sqrt{(n+1)(n+\ell+1)}\phi_{n+1}^{\ell}(u),
\end{eqnarray}
where $\tilde \varpi=2-\varpi$.
For the sake of simplicity we define the raising operator $K_+$ and the lowering operator $K_-$ acting on the wave function $ \phi_n^{\ell}(u,t)$ as
\begin{eqnarray}  \label{eq3.8}
K_-=\left(-u\frac{d}{du}+\frac{\ell}{2}+n-\frac{\varpi}{2}u\right),\\
K_+=\left(u\frac{d}{du}+\frac{\ell}{2}+n-\frac{\tilde \varpi}{2}u+1\right),
\end{eqnarray}
 and  hence obtain
 \begin{eqnarray}  \label{eq3.9}
  K_-\phi_n^{\ell}(u,t)&=&\sqrt{n(n+\ell)}\phi_{n-1}^{\ell}(u,t),\\
  K_+\phi_n^{\ell}(u,t)&=&\sqrt{(n+1)(n+\ell+1)}\phi_{n+1}^{\ell}(u,t).
 \end{eqnarray}
By multiplying  both side of the latter  equations  by the factor $e^{i\gamma_n^\ell (t)}$  we obtain
\begin{eqnarray}  \label{eq3.10} %\label{s}
 K_-\psi_n^{\ell}(u,t)&=& \sqrt{n(n+\ell)}\psi_{n-1}^{\ell}(u,t),\\ \label{eq3.11}
  K_+\psi_n^{\ell}(u,t)&=&\sqrt{(n+1)(n+\ell+1)}\psi_{n+1}^{\ell}(u,t).
\end{eqnarray}
 By successively applying $K_+$ on the ground state $\psi_0^{\ell} (u)$, we generate the eigenfunction  $\psi_n^{\ell} (u,t)$ of the system 
 as follows
 \begin{eqnarray} \label{eq3.12}
  \psi_n^{\ell}(u,t)&=&\sqrt{\frac{\Gamma(1+\ell)}{n!\Gamma(n+\ell+1)}}(K_+)^n\psi_0^{\ell} (u,t),\\
 \end{eqnarray}
 where,
 \begin{eqnarray} \label{eq3.13}
 \psi_0^{\ell}(u,t)&=&\frac{N(\rho,\theta)}{\sqrt{\Gamma(\ell+1)}}u^{\frac{\ell}{2}}e^{-\frac{\varpi}{2}u}e^{i\gamma_n^\ell(t)},\\ \label{eq3.14}
 K_- \psi_0^{\ell}(u,t)&=&0.
 \end{eqnarray}
 One can also observe that the following relations are satisfied
\begin{eqnarray} \label{eq3.15}
 K_+K_-\psi_n^{\ell} (u,t)&=&n(n+\ell)\psi_n^{\ell} (u,t),\\ \label{eq3.16}
 K_-K_+ \psi_n^{\ell} (u,t)&=&(n+1)(n+\ell+1)\psi_n^{\ell} (u,t).
\end{eqnarray}
Now, to establish the dynamical Lie algebra associated with the ladder operators $K_\pm$, we calculate the
commutator
\begin{equation} \label{eq3.17}
 [K_-,K_+]\psi_n^{\ell}(u,t)=(2n+\ell+1)\psi_n^{\ell}(u,t).
\end{equation}
As a consequence, we can introduce the operator $K_0$ defined to satisfy
\begin{eqnarray} \label{eq3.19}
 K_0\psi_n^{\ell}(u,t)=\frac{1}{2}(2n+\ell+1)\psi_n^{\ell}(u,t).
\end{eqnarray}
The operators $K_\pm$ and $K_0$  satisfy the following  commutation relations
\begin{eqnarray} \label{eq3.20}
 [K_-,K_+]=2K_0,\,\,[K_0,K_\pm]=\pm K_\pm,
\end{eqnarray}
which can be recognized as commutation relation of the generators of a non-compact  and non-abelian $SU(1,1)$ Lie group.
The corresponding Casimir operator for any irreducible representation is the identity times a number
\begin{eqnarray} \label{eq3.21}
 K^2=K_0^2-\frac{1}{2}(K_+K_-+K_-K_+)=\frac{1}{4}(\ell+1)(\ell-1).
\end{eqnarray}
It satisfies
\begin{equation} \label{eq3.22}
 [K^2,K_\pm]=0=[K^2,K_0].
\end{equation}
If we make the following   connection between  the physical quantum numbers $(n,\ell)$ and the  ordinary   $su(1,1)$   group numbers $(n,k)$ such as               
\begin{eqnarray}
 \ell=2k-1,   
\end{eqnarray}
  then we recover the ordinary discrete representations of the $su(1,1)$ Lie algebra 
  \begin{eqnarray}
   K^2\psi_n^{k}(u)&=&k(k-1)\psi_n^{k}(u),\\
   K_-\psi_n^{k}(u)&=& \sqrt{n(n+2k-1)}\psi_{n-1}^{k}(u),\label{s}\\
  K_+\psi_n^{k}(u)&=&\sqrt{(n+1)(n+2k)}\psi_{n+1}^{k}(u),\\   
  K_0\psi_n^{k}(u)&=& (n+k)\psi_n^{k}(u).
  \end{eqnarray}
Thus, in what   follows   we use the Bargmann index $\ell$ instead of the ordinary  index $k$ in  the  representation of $su(1,1)$ algebra. 
Now, with the properties of the generators $K_\pm$ and $K_0$ of this  algebra, we are in the position to construct the corresponding
coherent states  to this system.

\section{Barut-Girardello like coherent states} \label{section4}
\subsection{Contruction}

Following the Barut and Girardello approach \cite{10},  $ SU(1,1)$ coherent states  are defined  to  be 
the eigenstates of the lowering generator $K_-$
\begin{equation}\label{eq4.1} 
 K_-|\psi_z^\ell (t)\rangle=z |\psi_z^\ell (t)\rangle,
\end{equation}
where $z$ is an arbitrary complex number. The normalized Barut-Girardello states can be decomposed
over the number-state basis $|\psi_n^{\ell}(t)\rangle$ as follows 
\begin{eqnarray} \label{eq4.2} % \label{cbar}
 |\psi_z^\ell (t)\rangle&=&
 \sqrt{\frac{|z|^{\ell}}{I_{\ell}(2|z|)}}\sum_{n=0}^\infty\frac{z^n}{\sqrt{n!\Gamma(n+\ell+1)}}|\psi_n^{\ell}(t)\rangle,\\\label{eq4.2a}
 \psi_z^{\ell}(u,t)&=&\frac{|z|^{\frac{\ell}{2}} N_n(\rho,\alpha)}{\sqrt{I_{\ell}(2|z|)}}
\sum_{n=0}^\infty \frac{z^n}{\Gamma(n+\ell+1)}u^{\frac{\ell}{2}}e^{-\frac{\varpi}{2}u}L_n^{\ell}(u)e^{i\gamma_n^\ell (t)}.
\end{eqnarray}
However, in term of the generating function  (\ref{eq3.2}), the Barut-Girardello coherent  states can be written as follows
\begin{eqnarray} \label{eq4.3}
  \psi_z^{\ell}(u,t)&=&\left(\frac{z}{|z|}\right)^{-\frac{\ell}{2}}\frac{ N_n(\rho,\alpha) e^{z-\frac{\varpi}{2}u} }{\sqrt{I_{\ell}(2|z|)}}
 J_{\ell}\left(2\sqrt{uz}\right)e^{i\gamma_n^\ell (t)}.
\end{eqnarray}
As it is seen from Eq.(\ref{eq4.2}), the $|\psi_z^\ell (t)\rangle$ is a linear combination of number states $|\psi_n^\ell(t)\rangle$.
Therefore, the Barut-Giraldello like coherent states belong to
   Hilbert space $\mathcal{H}^\ell$ indexed by   
the single real positive number $\ell$ which defines  the   representation of this space
\begin{eqnarray} \label{eq4.4}
 \mathcal{H}^\ell:=\textit{span}\huge\{|\psi_n^\ell(t)\rangle\huge\}_{n\in\mathbb{N}}^\ell.
\end{eqnarray}

\subsection{The mathematical properties}
In the following discussion we will consider various properties of these states including the  non-orthogonality, the usual conditions of continuity in the label, normalizability,
 the resolution of identity by finding the weight function $\omega^\ell$ \cite{37'} . 
\subsubsection{\textit{The non-orthogonality}}
We can see that the scalar product of two  coherent states does not vanish
\begin{equation} \label{eq4.5}
 \langle\psi_{z_1}^\ell(t)|\psi_{z_2}^\ell(t)\rangle=\frac{I_{\ell}(2\sqrt{ z_1^*z_2})}{\sqrt{I_{\ell}(2|z_1|)I_{\ell}(2|z_2|)}}.
\end{equation}
 In the case $z_1 = z_2 = z$, we obtain the normalization
\begin{equation} \label{eq4.6}
 \langle\psi_{z}^\ell (t)|\psi_{z}^\ell(t)\rangle=1.
\end{equation}
 Thus, as it is well-known,  the states (\ref{eq4.2}) are normalized but  are not mutually  orthogonal.
\subsubsection{\textit{The Label continuity}}
The  continuity in label $z$  can then be stated as
\begin{eqnarray} \label{eq4.7}
 |||\psi_z^\ell (t)\rangle-|\psi_{z'}^\ell(t)\rangle||^2&=&2\left[1-\mathcal{R}e\left(\langle\psi_{z'}^\ell (t)|\psi_{z}^\ell(t)\rangle\right)\right]
 \longrightarrow 0,\cr&&   \mbox{when} \quad   |z-z'|^2 \rightarrow 0.
\end{eqnarray}

\subsubsection{\textit{Resolution of unity}}

\noindent The overcompleteness relation reads  as follows
\begin{equation} \label{eq4.8}
 \int d\mu(z,\ell)|\psi_z^\ell (t)\rangle\langle \psi_z^\ell(t)|=\sum_{n=0}^\infty|\psi_n^{\ell}(t)\rangle\langle \psi_n^{\ell}(t)|=
  \mathbb{I}^\ell,
\end{equation}
with the measure
\begin{equation} \label{eq4.9}
 d\mu(z,\ell)=\frac{2}{\pi}K_{\ell}(2|z|)I_{\ell}(2|z|)d^2z,
\end{equation}
where $d^2z=d(Re z)d(Im z)$ and $K_\upsilon(x)$ is the $\upsilon$-order modified Bessel function of the second kind.
 The weight-function of these BGLCSs are  then given by 
\begin{equation} \label{eq4.9bis}
\omega^\ell(|z|) = \frac{2}{\pi}K_{\ell}(2|z|)I_{\ell}(2|z|) \quad \ell = 0.5, 1, 1.5, \ldots
\end{equation}
The resolution of this identity is easy to demonstrate by using the following integral \cite{41}
\begin{eqnarray} \label{eq4.10}
 \int_0^\infty dx x^\mu K_\upsilon (ax)=2^{\mu-1} a^{-\mu-1}\Gamma \left(\frac{1+\mu+\upsilon}{2}\right)
 \Gamma \left(\frac{1+\mu-\upsilon}{2}\right),
\end{eqnarray}
where $[\mathcal{R}e(\mu+1\pm \upsilon)>0, \mathcal{R}e (a)>0 $] and 
all the integrals are performed over the whole complex $z$ plane, where $z=re^{i\varphi}$ \,\, $r\in[0,\infty[$ and 
$\varphi\in[0,2\pi]$.
 The asymptotic expression of the weight function $\omega^\ell(|z|)$ for $|z| \gg 1$ is, 
\begin{equation} \label{eq4.10a}
\omega^\ell(|z|) \simeq   {1\over 2\pi |z|}\left(1 + {1\over |z| } \left({\ell^2 \over 2} - {1 \over 8} \right)\right) 
%\quad \textrm{for} \ |z| \gg 1
\end{equation}
as the behaviours of the Bessels functions for $|z| \gg 1$ are 
\begin{eqnarray*}
K_\mu(x) & \simeq &  e^{-x}\sqrt{\pi \over 2 x}\left(1 + {1\over x } \left({\mu^2 \over 2} 
- {1 \over 8} \right) \right) \\
I_\mu(x) & \simeq &  {e^{x} \over \sqrt{2 \pi x}}. 
\end{eqnarray*}

\begin{figure}[htbp]
\resizebox{0.82\textwidth}{!}{
\includegraphics{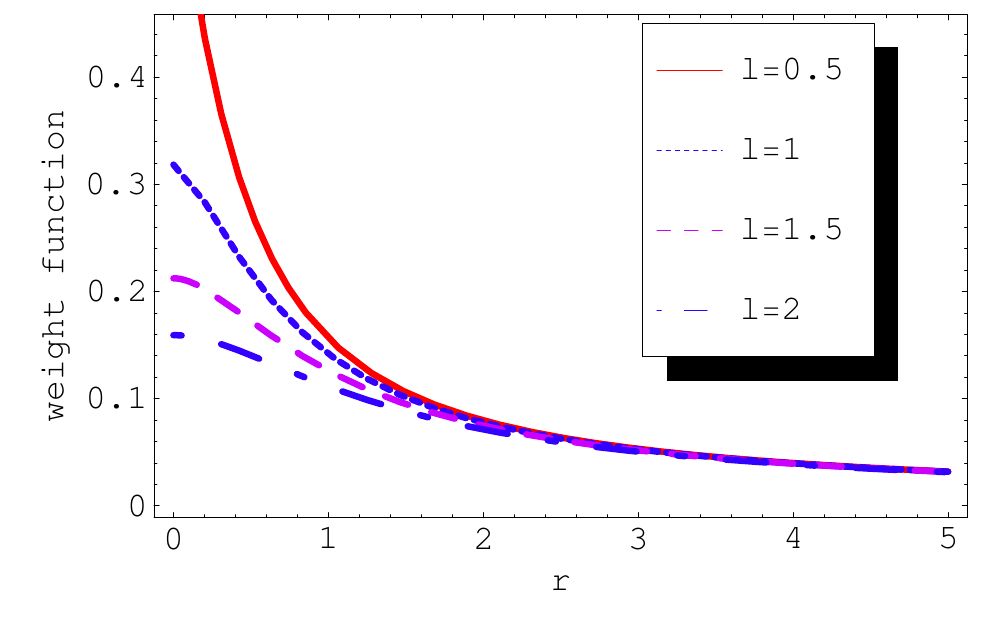}
}
\caption{\it \small 
{  Plots of the  weight function (\ref{eq4.9}) of  the
BGLCSs  (\ref{eq4.2}) versus  $ r = |z|$  for different  values of 
 the bargmann index $\ell$.}
}
\label{weight_CS}       
\end{figure}

Since the measure in equation (\ref{eq4.9}) must be necessary positive, the
function $\omega^\ell(|z|)$ must be  positive.  This is confirmed in Fig.\ref{weight_CS}    where  we  represent  the weight functions  (\ref{eq4.9bis}) for different values of the Bargmann index $\ell,\, \{ \ell = 0.5, 1, 1.5, 2\}$.
 We observe that  the weight function  globally decreases  and tends to $0$ when $|z|$ increases,  as confirmed by the asymptotic expressions (\ref{eq4.10a}). We can see also that the weight  function decreases while the Bargmann index $\ell$ increases.

 \subsection{The statistical  properties }
 After mathematical construction of the  BGLCSs, in the present subsection, we
investigate some of the quantum statistical properties of these states, such as the photon-number distribution,
the mean number of photons, the intensity correlation   function, the Mandel parameter and the Wigner function.
 \subsubsection{\textit{The photon-number distribution}}
\begin{figure}[htbp]
\resizebox{0.55\textwidth}{!}{
\includegraphics{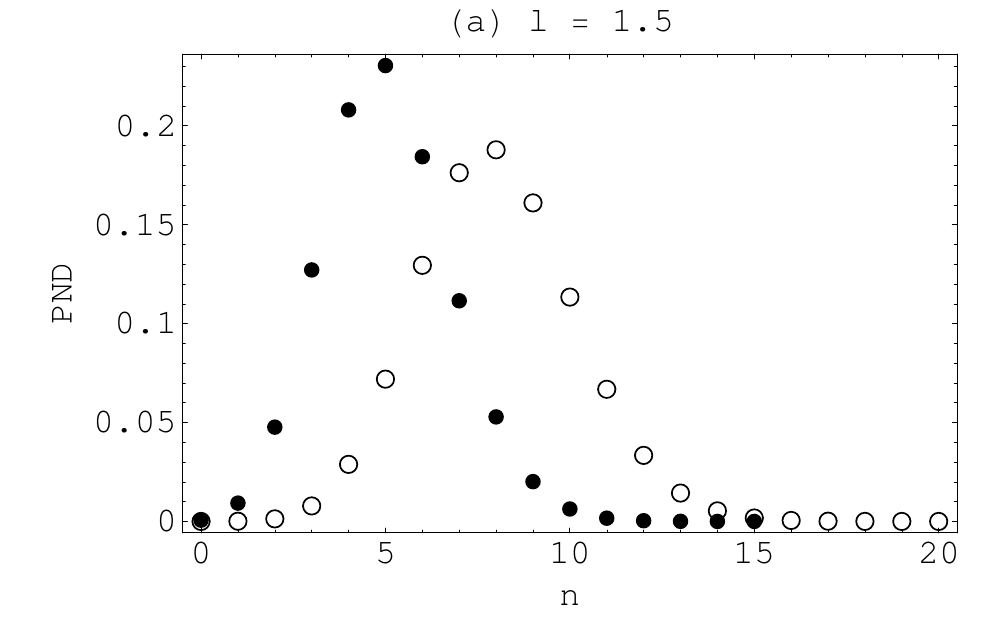}
}
\resizebox{0.55\textwidth}{!}{
\includegraphics{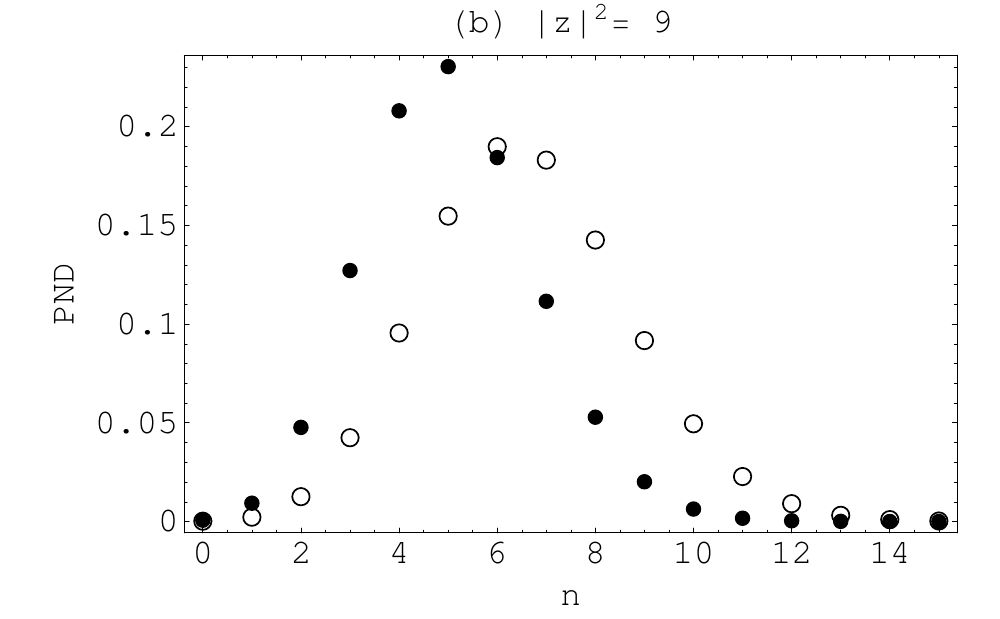}
}
\caption{\it \small 
 Plots of the PND (\ref{eq4.14}) of   the BGLCSs  (\ref{eq4.2}) versus the photon number $n$, 
with  parameters : (a)  fixed value of $\ell = 1.5$ with different values of the amplitude 
 $ |z|^2 = 6$  (dot points) and  $ |z|^2 = 9$ (scattered points);  (b)  fixed value of the amplitude  
 $|z|^2 = 9$  with different values of the Bargmann index $\ell = 3.5$  (dot points) and $\ell = 5$ (scattered points).}
\label{PND-CS}       
\end{figure}

 The probability of finding the  $n^{th}$ photons
in the states $|\psi_z^\ell(t)\rangle$ is given by
 \begin{eqnarray} \label{eq4.14}
  P_n(\ell,|z|)=|\langle\psi_{n}^\ell(t)|\psi_{z}^\ell(t)\rangle|^2=\frac{|z|^{2n+\ell}}{I_\ell(2|z|)n!\Gamma(n+\ell+1)}.
 \end{eqnarray}
 The complexity of the above equation makes it difficult to predict  analytically the statistical nature of these states. Therefore,
for two  limiting cases of the variable $|z|$ ($|z|\ll1$ and $|z|\gg 1$), the modified Bessel function $I_\mu(x)$ is respectively
approximated as follows \cite{37}
\begin{eqnarray} \label{eq4.15}%\label{f3}
 I_\mu(x)\simeq \frac{1}{\Gamma(\mu+1)}\left(\frac{x}{2}\right)^\mu \quad \mbox{and}\quad
 I_\mu(x)\simeq\frac{e^x}{\sqrt{2\pi x}}\left[1+O\left(\frac{1}{x}\right)\right].
\end{eqnarray}
Using these relations, the asymptoptical expressions of  the photon-number distribution of the  BGLCSs  (\ref{eq4.2}) are 
 \begin{eqnarray} \label{eq4.16}
 \lim_{|z|\rightarrow 0} P_n(\ell,|z|)=\frac{|z|^{2n}}{n!}\frac{\Gamma(\ell+1)}{\Gamma(n+\ell+1)},
 \quad \mbox{and}\quad 
 \lim_{|z|\rightarrow +\infty} P_n(\ell,|z|)=0.
 \end{eqnarray}
  So, for   small values of $|z|$ this distribution  is smaller than unity. The corresponding BGLCSs have sub-Poissonian statistics,
  while for large $|z|$, the probabilities $P_n(\ell,|z|)$ is zero.  
 In Figure \ref{PND-CS}, we plot  the  PND   as a function of the photon number $n$ for different parameters:  (a) fixed  Bargmann index  
$\ell =1.5$  and different values of $|z|^2 = \{6, 9\}$; (b) fixed value of $|z|^2 = 9$ and different values of the Bargmann index  $\ell =\{3.5, 5\}$.
We see in Figure (a) that as $z$ increases, the peaks decrease and shift to the right. Figure (b) shows that the shift in the PND is
less accentuated  as the Bargmann index $\ell$ increases.
\\
\subsubsection{\textit{The intensity correlation   function, the Mandel parameter and the Wigner function}}
\begin{figure}[htbp]
\resizebox{0.55\textwidth}{!}{
\includegraphics{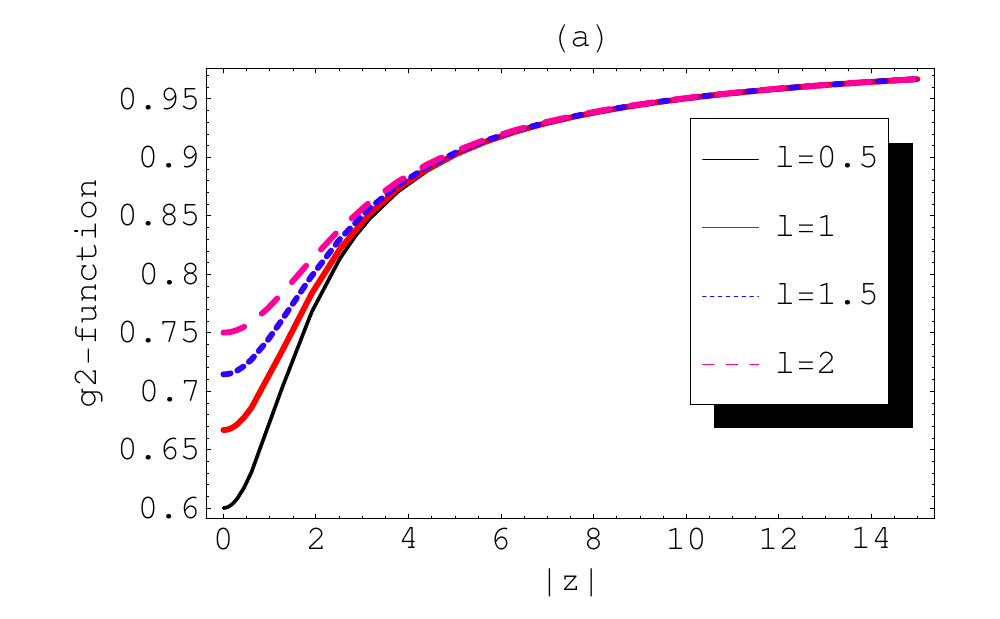}
}
\resizebox{0.55\textwidth}{!}{
\includegraphics{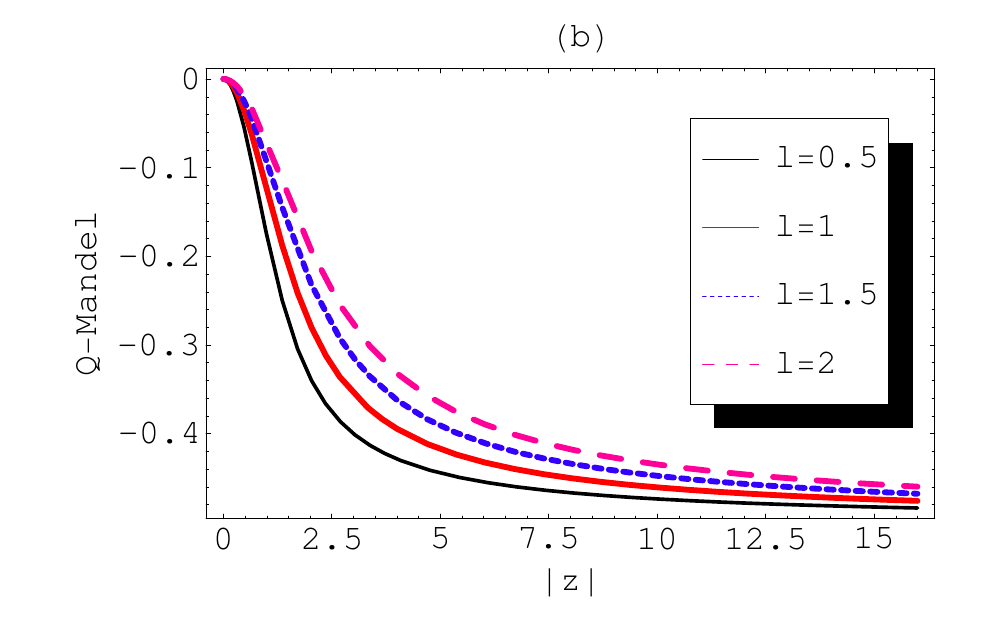}
}
\caption{\it \small 
Plots of the  intensity correlation function  (\ref{eq4.21}) (a)  and the Mandel Q-parameter  (\ref{eq4.23}) versus  $ |z|$  for different  values of 
 the bargmann index $\ell$.
}
\label{g2_QMandel}       
\end{figure}

 The intensity correlation function or equivalently  the Mandel Q-parameter 
yields the information about photon statistics of the quantum states. 
The intensity correlation function  of the BGLCSs (\ref{eq4.2}) is  defined by
\begin{eqnarray} \label{eq4.17}
 g_{\ell z}^{(2)}&=&\frac{\langle N^2\rangle_{z}^\ell-\langle N\rangle_{z}^\ell}{[\langle N\rangle_z^\ell]^2} ,
\end{eqnarray}
where $N$ is the number operator which is   defined as the operator which diagonalizes
the basis for the number states :
\begin{equation} \label{eq4.18}
 N |\psi_n^\ell(t)\rangle=n|\psi_n^\ell(t)\rangle.
\end{equation}
The  Mandel Q-parameter  is related to the intensity correlation function  by
\begin{eqnarray} \label{eq4.19}
 \mathcal{Q}_z^\ell &=& \langle N\rangle_{z}^\ell \left[g_{\ell z}^{(2)}-1\right].
\end{eqnarray} 
 The intensity correlation function (or the Mandel Q-parameter) determines whether  the 
BGLCSs  have a  photon number distribution.
This latter is sub-Poissonian  if  $g^2 <  1$  (or $ -1 \le Q < 0$), Poissonian if 
 $g^2 =  1$  (or $Q = 0$), and 
super-Poissonian  if  $g^2 > 1$ (or $Q > 0$). \\\\
We check that, for  BGLCSs (\ref{eq4.2}), the expectation values of $N$ and $N^2$ can be computed as \cite{34}
\begin{eqnarray} \label{eq4.20}
\langle N\rangle_{z}^\ell & = &  \langle \psi_z^\ell(t)| N|\psi_z^\ell (t)\rangle =|z|\frac{I_{\ell+1}(2|z|)}{I_\ell(2|z|)},\\ \label{eq4.20a}
\langle N^2\rangle_{z}^\ell & =  &  \langle \psi_z^\ell(t)| N^2|\psi_z^\ell (t)\rangle =|z|\frac{I_{\ell+1}(2|z|)}{I_\ell(2|z|)}+
 |z|^2\frac{I_{\ell+2}(2|z|)}{I_\ell(2|z|)} \label{eq4.20b}.
 \end{eqnarray}
 Taking into account  the results (\ref{eq4.20}, \ref{eq4.20a}) of the expectation values of the number operator and its square, we obtain
\begin{eqnarray} \label{eq4.21}
 g_{\ell z}^{(2)}&=& \frac{I_{\ell+2}(2|z|)I_\ell(2|z|)}{[I_{\ell+1}(2|z|)]^2}.
\end{eqnarray}
Using the above approximation conditions (\ref{eq4.15}) in case of $|z|\ll1$ and $|z|\gg 1$, the intensity correlation function are given by
\begin{eqnarray} \label{eq4.22}
  g_{\ell z}^{(2)}&\simeq&\frac{\ell+1}{\ell+2} \quad \mbox{for}\quad |z|\ll1,\\
   g_{\ell z}^{(2)}&\simeq& 1    \quad \mbox{for}\quad  |z|\gg 1.
\end{eqnarray}
For all $\ell$ values, the  BGLCSs  have sub-Poissonian statistics for small  values of $|z|$, while for large $|z|$, these states tend to have Poissonian statistics.\\\\
\noindent The  Mandel parameter $\mathcal{Q}_z^\ell$ is given by 
\begin{eqnarray} \label{eq4.23}
 \mathcal{Q}_z^\ell &=& |z|\frac{I_{\ell+2}(2|z|)I_{\ell}(2|z|)-I_{\ell+1}^2(2|z|)}{I_{\ell+1}(2|z|)I_{\ell}(2|z|)},\\ \label{eq4.23a}
        \mathcal{Q}_z^\ell      &\simeq& - \frac{|z|^2}{(\ell+1)(\ell+2)}  \quad \mbox{for}\quad |z|\ll1.\label{eq4.23b}
\end{eqnarray}
 In Figure \ref{g2_QMandel}, the intensity correlation function and the Mandel Q-parameter
have been plotted in terms of the amplitude $|z|$, in (a) and (b), respectively for different values of the Bargmann index $\ell = \{0.5, 1, 1.5 , 2\}$.
We can see that the  Mandel Q-parameter is always negative and the intensity correlation function  
$ g_{\ell z}^{(2)} <  1$ that confirms the analytical  forms 
(\ref{eq4.23}, \ref{eq4.23b}) and  showing that  the BGLCSs (\ref{eq4.2}) 
have sub-Poissonian  statistics.

To analyze the behavior on phase space $(u,p)\in\mathbb{R}^2$ of the Barut-Girardello like coherent states associated with this system, we  use
the Wigner  quasiprobability distribution function \cite{42} 
\begin{eqnarray} \label{eq4.24}% \label{f4}
 \mathcal{W}_z^\ell(u, p,t)&=&\frac{1}{2\pi\hbar}\int_{-\infty}^{+\infty}e^{-i\frac{pv}{\hbar}}{\psi_{z'}^*}^{\ell}\left(u-\frac{v}{2},t\right)
{\psi_z}^{\ell}\left(u+\frac{v}{2},t\right)dv,\cr
&=& \frac{|z|^{\ell}}{2\pi \hbar I_{\ell}(2|z|)} \sum_{n'n=0}^\infty\frac{z^n {z^*}^{n'}}
{\sqrt{n!n'!\Gamma(n+\ell+1)\Gamma(n'+\ell+1)}}\cr&&\times
\int_{-\infty}^{+\infty}e^{-i\frac{pv}{\hbar}}{\psi_{n'}^*}^{\ell}\left(u-\frac{v}{2},t\right)
{\psi_n}^{\ell}\left(u+\frac{v}{2},t\right)dv.
\end{eqnarray}
In order to compute  the  integral of the wave functions $ \psi_n^\ell(u,t)$ in (\ref{eq4.24})
it is convenient to use together the following change of variables $r=e^{-\frac{\varpi}{2}v}$ and $\xi= e^{-\varpi u}$ \cite{42,43}, 
so 
\begin{eqnarray} \label{eq4.25}
 \mathcal{W}_z^\ell(u, p,t)&=&\frac{|z|^\ell }{\pi\varpi\hbar I_\ell(2|z|)}\xi^\ell\sum_{n'n=0}^\infty
 N_n(\rho,\theta)N_{n'}(\rho,\theta)\cr&&\times
  \frac{z^n {z^*}^{n'}e^{i(\gamma_n^\ell(t)-\gamma_{n'}^\ell(t))}} {\Gamma(n+\ell+1)\Gamma(n'+\ell+1)}\cr&&\times
  \int_0^{\infty}e^{-\frac{\varpi \xi}{2}(r+r^{-1})}L_{n'}^\ell(\xi r^{-1})L_{n}^\ell(\xi r)r^{\frac{-2ip}{\hbar\varpi}-1}dr.
\end{eqnarray}
Then,  expanding  the  associated  Laguerre  polynomials  in
their finite series with binomial coefficients (\ref{eq3.1}) and using the following relation \cite{31}
\begin{eqnarray} \label{eq4.26}
 \int_0^\infty dx x^{\nu-1} e^{-\frac{\lambda}{x}-\tau x}=2\left(\frac{\lambda}{\tau}\right)^{\frac{\nu}{2}}K_\lambda (2\sqrt{\lambda\tau}),
\end{eqnarray}
we get the final result 
\begin{eqnarray} \label{eq4.27}
\mathcal{W}_z^\ell(u, p,t)&=&2\frac{|z|^\ell }{\pi\varpi\hbar I_\ell(2|z|)}\xi^\ell\sum_{n'n=0}^\infty N_n(\rho,\theta)N_{n'}(\rho,\theta)\cr&&\times
  \frac{z^n {z^*}^{n'}e^{i(\gamma_n^\ell(t)-\gamma_{n'}^\ell(t))}} {\Gamma(n+\ell+1)\Gamma(n'+\ell+1)}\cr&&
  \sum_{k'=0}^{n'} \sum_{k=0}^{n}\binom{n'+\ell}{n'-k'}\binom{n+\ell}{n-k}\frac{(-\xi)^{k+k'}}{k'!k!}
  K_{k-k'-\frac{2ip}{\hbar\varpi}}\left(\varpi \xi\right),
\end{eqnarray}
where $K_\lambda(\xi)$ are the modified Bessel functions of the third kind. This result is related  to the similar discussions of the 
Wigner function of Morse potential \cite{42}.

\section{ The photon added   coherent states }\label{section5}
\subsection{Construction}
 Photon-added   coherent states 
$|\psi_{zm}^\ell(t)\rangle$ are defined by the repeated application  of the raising   operator $K_+$
 to the BGLCSs (\ref{eq4.2})  of the TDLP
 \begin{eqnarray} \label{eq5.1}
  |\psi_{zm}^\ell(t)\rangle&=&{(K_+)^m|\psi_z^\ell (t)\rangle\over \sqrt{\langle \psi_{z}^\ell (t)|(K_-)^m(K_+)^m|\psi_{z}^\ell (t)\rangle }}\cr
                        &=& \mathcal{M}_{ m}^\ell (|z|)                        \sum_{n=0}^\infty\frac{z^n}{\sqrt{n!\Gamma(n+\ell+1)}} (K_+)^m|\psi_n^{\ell}(t)\rangle
\end{eqnarray}
where $m$ is a positive integer being the number of added  quanta (or added photons). 
$\mathcal{M}_{m}^\ell(|z|)$ is the normalization constant such as
\begin{eqnarray} \label{eq5.2}
 \mathcal{M}_{ m}^\ell (|z|)&=&\frac{\sqrt{\frac{|z|^{\ell}}{I_{\ell}(2|z|)}}}{\sqrt{\langle \psi_{z}^\ell(t)|(K_-)^m(K_+)^m|\psi_{z}^\ell(t)\rangle }}.
\end{eqnarray}
 For $m=0$, we recover the normalization constante of the  states (\ref{eq4.2a}).
 Making use of the expressions
\begin{eqnarray}  \label{eq5.3}
  (K_+)^m|\psi_n^{\ell}(t)\rangle & = & \sqrt{\frac{\Gamma(n+m+1)\Gamma(n+\ell+m+1)}{\Gamma(n+1)\Gamma(n+\ell+1)}}|\psi_{n+m}^{\ell}(t)\rangle,\label{eq5.31}
\end{eqnarray}
we  obtain
\begin{eqnarray}\label{eq5.4} \label{f5}
 |\psi_{zm}^\ell(t)\rangle_m=\mathcal{M}_{ m}^\ell (|z|)\sum_{n=0}^\infty\frac{z^n}{\sqrt{F_{ m}(\ell,n)}}|\psi_{n+ m}^\ell(t)\rangle,
\end{eqnarray}
  and 
\begin{eqnarray} \label{eq5.5}
 F_{ m}(\ell,n)&=&\frac{[\Gamma(n+\ell+1)]^2[\Gamma(n+1)]^2}{\Gamma(n+m+\ell+1)\Gamma(n+m+1)}.
\end{eqnarray}
The corresponding   eigenfunctions of the  states (\ref{eq5.4})  are termed
\begin{eqnarray} \label{eq5.6}
 \psi_{zm}^\ell(u,t)&=&\mathcal{M}_{m}^\ell (|z|)u^{\frac{\ell}{2}}
 e^{-\frac{\varpi}{2}u}\sum_{n=0}^\infty N_{n+m}(\theta,\rho)
 \frac{\Gamma(n+m+1)}{\Gamma (n+1)\Gamma(n+\ell+1)}\cr
 && \times z^n L_{n+m}^\ell (u) e^{i\gamma_{n+m}^\ell(t)}. \label{eq5.61}
\end{eqnarray}
As it is seen from Eq.(\ref{eq5.4}), the $|\psi_{zm}^\ell(t)\rangle$ are linear combination of number states $|\psi_{n+ m}^\ell(t)\rangle$.
Therefore,  the  PABGLCSs   belong to %  $(n+m)$-dimensional 
the Hilbert space $\mathcal{H}_{ m}^\ell$ indexed by  
the single real positive number $\ell$ which defines  the   representation of these spaces
\begin{eqnarray} \label{eq5.7}
 \mathcal{H}_{m}^\ell  := \textit{span}\huge\{|\psi_{n+m}^\ell(t)\rangle\huge\}_{n\in\mathbb{N}}^\ell .
\end{eqnarray}
In other words, the application of $(K_+)^m$  transfers the coherent states of Barut-Giraldello from 
$\mathcal{H}^\ell$ to 
%the  total Hilbert  space
 $\mathcal{H}_{ m}^\ell$.
\subsection{The mathematical properties}
In this subsection we are interested to examine the change  of the mathematical properties of the states (\ref{f5})
compared to the previous properties of BGLCSs.
\subsubsection{\textit{The non-orthogonality}}
The states $|\psi_{zm}^\ell(t)\rangle$  of  this system must  be normalized but not orthogonal. The non-orthogonality is expressed as 
\begin{eqnarray} \label{eq5.8}
\langle \psi_{z'm'}^\ell(t) |\psi_{zm}^\ell(t)\rangle&=& \mathcal{M}_{ m'}^\ell (|z'|)\mathcal{M}_{ m}^\ell (|z|)
\sum_{n'=0}^\infty\sum_{n=0}^\infty
  \frac{z^n (z'^*)^{n'}}{\sqrt{F_{ m'}(\ell,n')F_{ m}(\ell,n)}}\cr&&\times\langle \psi_{n'+ m'}^\ell (t) |\psi_{n + m}^\ell (t)\rangle.
\end{eqnarray}
 In the case of the state $|\psi_{zm}^\ell(t)\rangle$, due to the orthogonality relation of the number vectors $|\psi_n^\ell(t)\rangle$, it follows that
\begin{eqnarray} \label{eq5.9}
  \langle \psi_{z'm'}^\ell (t) |\psi_{zm}^\ell (t)\rangle &=& \mathcal{M}_{m'}^\ell (|z'|)\mathcal{M}_{m}^\ell (|z|)(z'^*)^{m-m'}
  \cr&&\times\frac{\Gamma(m+1)\Gamma(m+\ell+1)}
  {\Gamma(m-m'+1)\Gamma(m-m'+\ell+1)\Gamma(\ell+1)}\cr&&\times
  {}_2\!F_3 (m+1,m+\ell+1;\cr&& m-m'+1,m-m'+\ell+1,\ell+1;z'^*z),
\end{eqnarray}
where ${}_2\!F_3 $ is the generalized hypergeometric function and  with $m>m'$. This relation  can be obtained in a more explicit
way in terms of Meijer’s G-function by
\begin{eqnarray} \label{eq5.10}% \label{eq82}
\langle \psi_{z'm'}^\ell (t) |\psi_{zm}^\ell (t)\rangle&=& \mathcal{M}_{m'}^\ell (|z'|)\mathcal{M}_{m}^\ell (|z|)(z'^*)^{m-m'}\cr&&
 G_{2,4}^{1,2}
 \left( -z'^*z\big|_{0,\quad m'-m,\quad m'-m-\ell,\quad -\ell}^{-m,\,\,\, -m-\ell}\right),
\end{eqnarray}
where we used the following relation between the generalized hypergeometric function and 
the Meijer’s G-function
\begin{eqnarray} \label{eq5.11} 
 {}_p\!F_q (a_1,\cdots,a_p;b_1,\cdots,b_q;x)=\frac{\prod_{j=1}^q\Gamma(b_j)}{\prod_{j=1}^p\Gamma(a_j)}
 G_{p,q+1}^{1,p}
 \left( -x\big|_{\,\,0,\quad \quad\quad 1-b_q}^{(1-a_p)\quad}\right).
\end{eqnarray}
  This  relation (\ref{eq5.10})  prove that the PAGBLCSs are
not mutually orthogonal.\\
\noindent Performing the normalization condition of these states,  such that
\begin{eqnarray} \label{eq5.14}
 \langle \psi_{zm}^\ell(t) |\psi_{zm}^\ell(t)\rangle = 1,
\end{eqnarray}
 we determine the  constants   $\mathcal{M}_{ m}^\ell (|z|)$
\begin{eqnarray} \label{eq5.15}
 \mathcal{M}_{m}^\ell (|z|)  =  \left[G_{2,4}^{1,2}
 \left( -|z|^2\big|_{0,\,\,\,\,\,\,\,\,\,\,\,\,\quad 0,\quad  -\ell,\quad -\ell}^{-m,\,\,\, -m-\ell}\right)\right]^{-\frac{1}{2}}\label{eq5.15a}
\end{eqnarray}
From these results,  for $m=0$ the normalization constant (\ref{eq5.15a}) becomes
\begin{eqnarray}
 \mathcal{M}_0^\ell (|z|) & = & \left[G_{2,4}^{1,2}
 \left( -|z|^2\big|_{0,\,\,\,\,\,\,\,\,\,\,\,\,\quad 0,\quad  -\ell,\quad -\ell}^{0,\quad\quad \,\,\,\,-\ell}\right)\right]^{-\frac{1}{2}}=
 \sqrt{\frac{|z|^{\ell}}{I_{\ell}(2|z|)}},\label{eq5.15a}
\end{eqnarray}
which is  the normalization constant of the states (\ref{eq4.2a}). Therefore, with this condition the non-orthogonality (\ref{eq5.10}) is reduced 
to the one obtained (\ref{eq4.5}) for the states $|\psi_z^\ell(t)\rangle$. All this argument  shows that the addition of photons changes the non-orthogonality
 of the BGLCSs. 
\subsubsection{\textit{The label continuity}}
The label continuity condition of the $|\psi_{zm}^\ell(t)\rangle$ can
then be stated as:
\begin{eqnarray} \label{eq5.18}
|| |\psi_{zm}^\ell(t)\rangle-|\psi_{z'm'}^\ell (t)\rangle ||^2&=&
 2\left[1-\mathcal{R}e\left(\langle\psi_{z'm'}^\ell (t)|\psi_{zm}^\ell(t)\rangle\right)\right]\rightarrow0 \quad 
 \\
 \mbox{when} &&|z-z'|\rightarrow 0\quad \mbox{and}\quad |m-m'|\rightarrow 0 \nonumber
\end{eqnarray}

\subsubsection{\textit{Overcompleteness}}

 We have to search for  non-negative weight functions
$W_{ m}^\ell(|z|)$ such that the overcompleteness or the resolution
of the identity
%The resolution of unity operator in this space is
\begin{eqnarray}\label{eq5.18} %\label{f7}
 \int_{\mathbb{C}}\frac{d^2z}{\pi} |\psi_{zm}^\ell(t)\rangle  W_{ m}^\ell(|z|)
   \langle \psi_{zm}^\ell(t) |
 =\mathbb{I}_{ m}^\ell
\end{eqnarray}
holds, where    
\begin{eqnarray} \label{eq5.19}
\mathbb{I}_{ m}^\ell = \sum_{n=0}^\infty
 |\psi_{n+ m}^\ell(t)\rangle  \langle \psi_{n+ m}^\ell (t)|. \label{eq5.191}
\end{eqnarray}

\begin{figure}[htbp]
\resizebox{0.55\textwidth}{!}{
\includegraphics{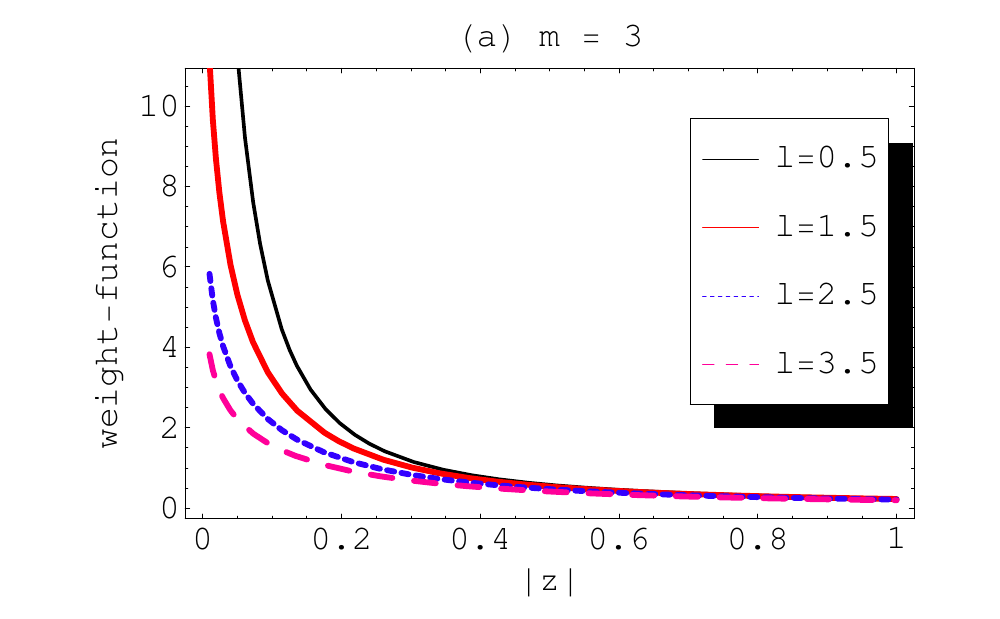}
}
\resizebox{0.55\textwidth}{!}{
\includegraphics{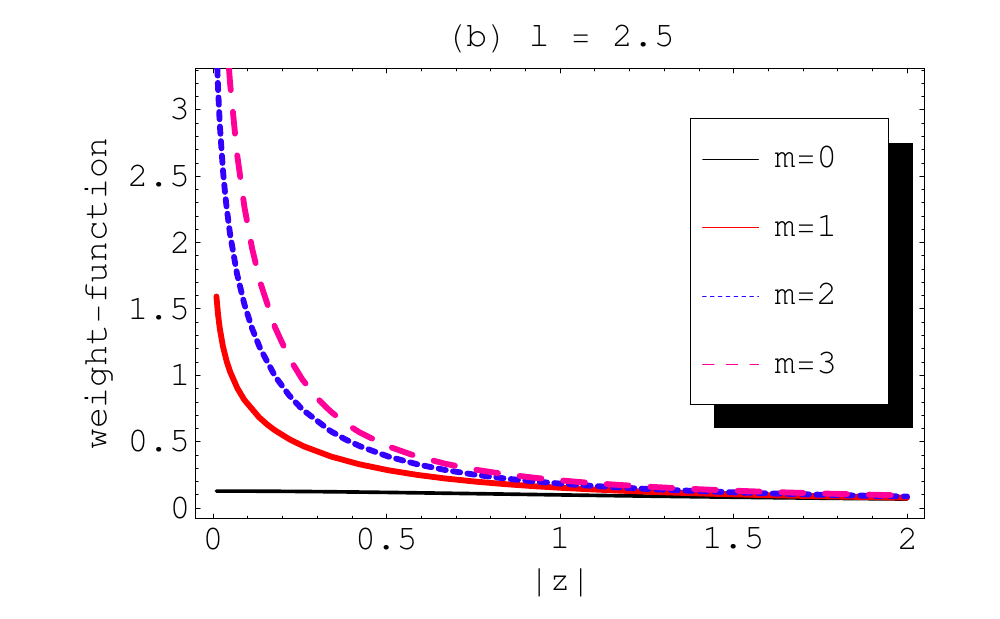}
}
\caption{\it \small 
 Plots of the  weight fuction $W_{m}^\ell$ (\ref{eq5.24}) of the PABGLCSs versus  $ |z|$:  
(a) for fixed value of the photon-added number $m = 3$ and different values of the  the Bargmann index $\ell$; 
(b) for fixed value of the  the Bargmann index $\ell = 2.5$ of the photon-added number $m = 3$ and different values of the photon-added number $m$.}
\label{weight-function-PACSDT}       
\end{figure}

For PABGLCSs case, by substituting equation (\ref{eq5.4}) into equation (\ref{eq5.18}) we obtain
\begin{eqnarray} \label{eq5.21}
 \int_{\mathbb{C}}\frac{d^2z}{\pi}W_{m}^\ell(|z|) [\mathcal{M}_{m}^\ell(|z|)]^2 \sum_{n,n'=0}^\infty 
 \frac{z^n (z^*)^{n'}}{\sqrt{F_{m}(\ell,n')F_{m}(\ell,n)}}
 \cr\times|\psi_{n+m}^\ell(t)\rangle  \langle \psi_{n+m}^\ell(t) |=\mathbb{I}_{m}^\ell.
\end{eqnarray}
By means of a change of the complex variables in terms of polar coordinates
$z=re^{i\vartheta}$ where $ r\in\mathbb{R}_+$, $\vartheta\in[0,2\pi)$, and  $d^2z=rdrd\vartheta$,
this  equation becomes
\begin{footnotesize}
\begin{eqnarray} \label{eq5.22}
 \sum_{n'n=0}^\infty\left[\frac{1}{\sqrt{F_{m}(\ell,n')F_{m}(\ell,n)}} \int_0^\infty dr r^{1+n+n'}
 W_{m}^\ell(|z|) [\mathcal{M}_{+m}^\ell (|z|)]^2\int_0^{2\pi}\frac{d\vartheta}{\pi}e^{i(n-n')\vartheta}\right]\cr\times
 |\psi_{n+m}^\ell(t)\rangle  \langle \psi_{n+m}^\ell(t) |=\mathbb{I}_{m}^\ell.
\end{eqnarray}
\end{footnotesize}
By performing the angular integration, i.e
\begin{eqnarray} \label{eq5.23}
 \int_0^{2\pi}\frac{d\vartheta}{\pi}e^{i(n-n')\vartheta}=2\delta_{nn'},
\end{eqnarray}
the resolution of the identity operator  is
\begin{eqnarray} \label{eq5.24}
 2\sum_{n=0}^\infty\left[\frac{1}{F_{m}(\ell,n)} \int_0^\infty dr r^{1+2n}
 W_{m}^\ell(r^2) [\mathcal{M}_m^\ell (r^2)]^2\right]\cr\times
 |\psi_{n+m}^\ell(t)\rangle  \langle \psi_{n+m}^\ell(t)|=\mathbb{I}_{m}^\ell.
\end{eqnarray}
Setting the weight function such as
\begin{eqnarray} \label{eq5.25} % \label{f10}
  W_{m}^\ell(r^2) =\frac{1}{[\mathcal{M}_{m}^\ell(r^2)]^2}r^{2m}g_m^\ell(r^2),
\end{eqnarray}  
and using the completeness of the states $|\psi_{n+m}^\ell(t)\rangle$ and  performing the variable change $x=r^2$ and $n+m=s-1$, 
the integral from the above equation is called the Mellin transform
\begin{eqnarray} \label{eq5.26} % \label{f8}
 \int_0^\infty dx x^{s-1}  g_m^\ell(x)=F_{m}(\ell,s-m-1)=\frac{[\Gamma(s+\ell-m)]^2[\Gamma(s-m)]^2}{\Gamma(s+\ell)\Gamma(s)}.
\end{eqnarray}
Using the definition of Meijers $G$-function, it follows that
\begin{eqnarray}\label{eq5.27} % \label{f9}
 \int_0^\infty dx x^{s-1}G_{p,q}^{m,n}
 \left(\alpha x\big|_{b_1,\cdots,\quad b_m,\quad b_{m+1},\quad \cdots, b_q}^{a_1,\cdots,\quad a_n,\quad a_{n+1},\quad \cdots, a_p}\right)\cr=
 \frac{1}{\alpha^s}
 \frac{\prod_{j=1}^m\Gamma(b_j+s)\prod_{j=1}^n\Gamma(1-a_j-s)}{\prod_{j=m+1}^q\Gamma(1-b_j-s)\prod_{j=n+1}^p\Gamma(a_j+s)}.
 \end{eqnarray}
 Comparing equations (\ref{eq5.26}) and (\ref{eq5.27}), we obtain that
 \begin{eqnarray} \label{eq5.28}
g_m^\ell(x)=G_{2,4}^{4,0}
 \left( x\big|_{-m,\quad -m,\quad \ell-m,\quad \ell-m}^{0,\,\,\,\,\quad\quad \ell}\right).
\end{eqnarray}
Using the multiplication formula of Meijer’s G-function
\begin{eqnarray} \label{eq5.29}
 x^\sigma G_{p,q}^{m,n}
 \left(\alpha x\big|_{(b_q)}^{(a_p)}\right)=G_{p,q}^{m,n}
 \left(\alpha x\big|_{(b_q+\sigma)}^{(a_p+\sigma)}\right),
\end{eqnarray}
then, the weight function (\ref{eq5.25}) becomes
\begin{eqnarray} \label{eq5.30}
 W_{m}^\ell(|z|)=\frac{1}{[\mathcal{M}_{m}^\ell (|z|)]^2 }G_{2,4}^{4,0}
 \left( |z|^2\big|_{0,\quad 0,\quad \ell,\quad \ell}^{m,\,\,\,\, m+ \ell}\right),
\end{eqnarray}
and  the overcompleteness (\ref{eq5.18}) can be explicitly expressed as follows
\begin{eqnarray}  \label{eq5.31}
 \int_{\mathbb{C}}\frac{d^2z}{\pi}\frac{1}{[\mathcal{M}_{m}^\ell (|z|)]^2  }G_{2,4}^{4,0}
 \left( |z|^2\big|_{0,\quad 0,\quad \ell,\quad \ell}^{m,\,\,\,\, m+ \ell}\right) |\psi_{zm}^\ell(t)\rangle\langle \psi_{zm}^\ell(t) |
 =\mathbb{I}_{m}^\ell.
\end{eqnarray}
Thus, at the limit $m=0$, this   weight function  is reduced to   the one obtained for the   ordinary Barut-Giraldello like coherent states (\ref{eq4.9bis})
\begin{eqnarray} \label{eq5.20}
\omega^\ell &=& \frac{1}{\pi} W_0^\ell(|z|)=\frac{1}{\pi} \left(\mathcal{M}_0^\ell(|z|)\right)^{-2}G_{2,4}^{4,0}\left( |z|^2\big|_{0,\quad 0,\quad \ell,\quad \ell}^{0,\,\,\,\, \ell}\right),\\
&=&\frac{2}{\pi}K_{\ell}(2|z|)I_{\ell}(2|z|).
\end{eqnarray}
   In Fig. \ref{weight-function-PACSDT}, we plot the weight function (\ref{eq5.30}) versus $x = |z|$ for fixed  
   value of phton-added number $m= 3$  and  various  values of the Bargmann index $\ell$ on figure (a)
   and fixed value of the parameter $\ell = 2.5$ and for  various values of $m$ on figure (b).  All the curves are positive, this confirms the positivity 
   of the weight function for half integer values of  the parameter $\ell = 0.5, 1.5, \ldots$.
All the curves on figure (a) have the same behaviour, this shows that the Bargmann index $\ell$ do not affect the general form of the curves. 
But we can see that the weight function decreases while $\ell$ increases. Figure (b) shows that increasing the photon-added number $m$ increases the weight function.  We see also  that all the curves tends asymptoptically to the weigt function of the conventional 
BGLCSs (\ref{eq4.9bis}).\\\\
\noindent 

\subsection{The statistical  properties}
We have  seen that the process of photon addition has  changed  the mathematical properties of the 
BGLCSs. Here, we  analyze   the effect of this process  on the statistics of the original coherent states.

\subsubsection{\textit{The photon-number distribution}}
\begin{figure}[htbp]
\resizebox{0.60\textwidth}{!}{
\includegraphics{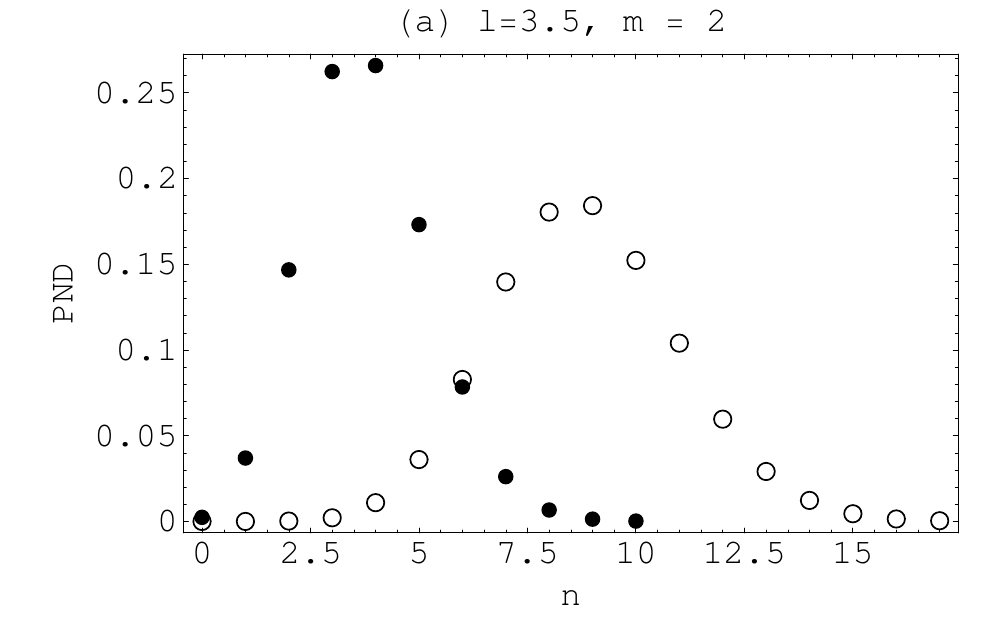}
}
\resizebox{0.60\textwidth}{!}{
\includegraphics{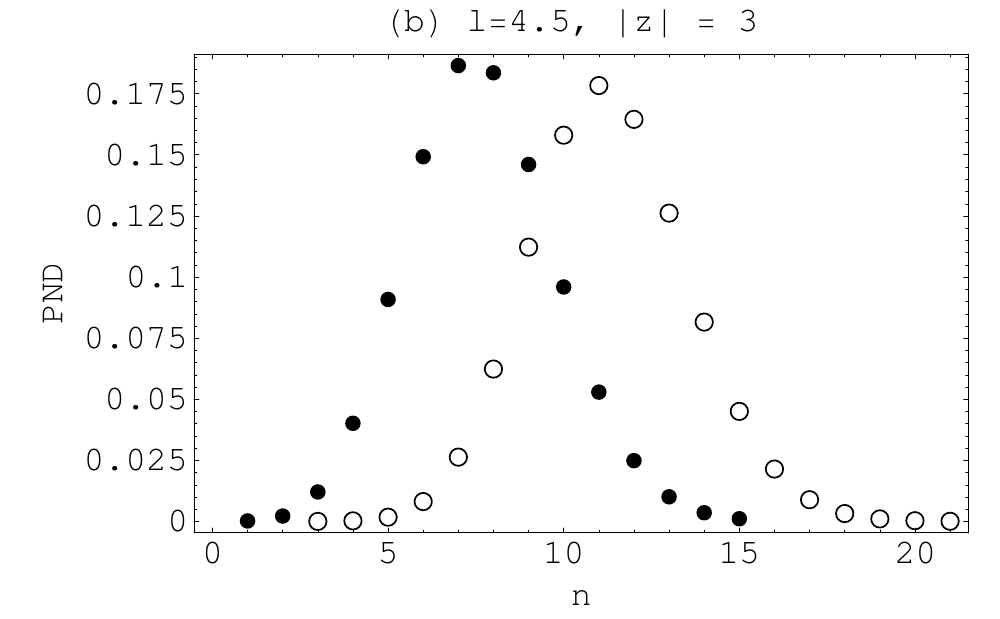}
}
\caption{\it \small 
{  Plots of the PND (\ref{eq5.40})   of the PABGLCSs   versus the photon number $n$, 
with  parameters : (a)  fixed value of $\ell = 3.5$ and $m = 2$ and  different values of the amplitude 
 $ |z|= 2$  (dot points) and  $ |z| = 3$ (scattered points);  (b)  fixed value of $\ell =4.5$ and  $|z| = 9$  with different values of  photon-added number $m = 1$  (dot points) and $m = 7$ (scattered points).}
}
\label{PND-PACS}       
\end{figure}

The  probability of finding $n$ photons in the  PAGBLCSs    in the Hilbert  spaces $\mathcal{H}_{ m}^\ell$ are  given by :
\begin{eqnarray} \label{eq5.40}
P_{n  m}(\ell,z)&=& |\langle \psi_{n}^\ell(t)|\psi_{zm}^\ell(t)\rangle|^2
               = [\mathcal{M}_{ m}^\ell(|z|)]^2}{|z|^{2n} \over  F_{ m}(\ell,n).            
\end{eqnarray}
The   addition of photons  drastically increases the photon distribution (\ref{eq4.14})  such  as 
\begin{eqnarray}
 P_{n + m}(\ell,z)&=& \frac{|z|^{2n}}{G_{2,4}^{1,2}
 \left( -|z|^2\big|_{0,\,\,\,\,\,\,\,\,\,\,\,\,\quad 0,\quad  -\ell,\quad -\ell}^{-m,\,\,\, -m-\ell}\right)}\cr&&\times
 \frac{\Gamma(n+m+\ell+1)\Gamma(n+m+1)}{[\Gamma(n+\ell+1)]^2[\Gamma(n+1)]^2}.
\end{eqnarray}
At the limit $m=0$, we recover the result of the photon distribution (\ref{eq4.14}) of the ordinary coherent states. This situation reflects the fact that $m$ photons have been added.  In Figure \ref{PND-PACS}, the photon number distribution of the PABGLCSs ( Figure.\ref{PND-PACS}), as function of $n$ is depicted in (a) with fixed parameters   
$\ell = 3.5$,  $m = 2$   different amplitude values  $|z| = \{4, 9\}$; in  (b) with fixed values of $\ell =4.5$,  $|z| = 9$ and different values of the photon-added number  $m =\{1, 7\}$.
We see in Figure (a) that as $|z|$ and $m$  increase, the peaks decrease and shift to the right. Figure (b) shows that the  shift of the peaks  increase as the photon-added number increase. 
\\\\

\subsubsection{\textit{The intensity correlation   function, the Mandel parameter and the Wigner function}}

The intensity correlation function $ g_{\ell, zm}^{(2)}$ and the Mandel $\mathcal{Q}_{zm}^\ell$-parameter are given by:
\begin{eqnarray} \label{eq5.46}
 g_{\ell, zm}^{(2)}&=&\frac{ \langle \psi_{zm}^\ell(t)|N^2|\psi_{zm}^\ell(t)\rangle - \langle \psi_{zm}^\ell|N|\psi_{zm}^\ell\rangle}
 {[ \langle \psi_{zm}^\ell|N|\psi_{zm}^\ell\rangle]^2},\\
\mathcal{Q}_{zm}^\ell&=&\langle \psi_{zm}^\ell(t)|N|\psi_{zm}^\ell(t)\rangle\left[g_{\ell}^{(2)}-1\right].
\end{eqnarray}

The expectation values of the operator number $N$ and its square in the states $ \psi_{n+ m}^\ell(t)$  are given by
\begin{eqnarray}  \label{eq5.42}
 \langle \psi_{n + m}^\ell(t)|N|\psi_{n+ m}^\ell(t)\rangle&=& n + m, \label{eq5.42 bis1}\\
 \langle \psi_{n + m}^\ell (t)|N^2|\psi_{n+ m}^\ell(t)\rangle &=& (n + m)^2.\label{eq5.42 bis2}
\end{eqnarray}
Based on the references \cite{32}, these  expectation values  in the $ |\psi_{zm}^\ell(t)\rangle$ states give 
\begin{footnotesize}
\begin{eqnarray}  \label{eq5.43} %\label{EV1}
 \langle \psi_{zm}^\ell(t)|N|\psi_{zm}^\ell(t)\rangle & = & 
 m - {\mathcal{G}_{ m}^\ell(1)\over \mathcal{G}_{ m}^\ell(0)},
  \\ \label{eq5.431} %\label{EV2}
% -\frac{G_{2,4}^{1,2} \left( -|z|^2\big|1\right)}
% {G_{2,4}^{1,2} \left( -|z|^2\big|0\right)},\\
  \langle \psi_{zm}^\ell(t)|N^2|\psi_{zm}^\ell(t)\rangle&=&
  m^2-(2m+1){\mathcal{G}_{ m}^\ell(1)\over \mathcal{G}_{ m}^\ell(0)} + {\mathcal{G}_{ m}^\ell(2)\over \mathcal{G}_{ m}^\ell(0)}, 
%  \frac{G_{2,4}^{1,2} \left( -|z|^2\big|1\right)}
% {G_{2,4}^{1,2} \left( -|z|^2\big|0\right)} %\cr&&
% +\frac{G_{2,4}^{1,2} \left( -|z|^2\big|2\right)}
% {G_{2,4}^{1,2} \left( -|z|^2\big|0\right)},
\end{eqnarray}
\end{footnotesize}
where 
\begin{eqnarray}
\mathcal{G}_{m}^\ell(i) 
 &=&G_{2,4}^{1,2}
 \left( -|z|^2\big|_{0,\quad\quad i,\quad\quad \,\,-\ell,\quad -\ell}^{-m,\,\,\, -m-\ell}\right),\qquad  i = 0, 1, 2.
\end{eqnarray}
For the special case $m=0$, we have 
\begin{eqnarray}
\mathcal{G}_{0}^\ell(0) 
 &=&G_{2,4}^{1,2}
 \left( -|z|^2\big|_{0,\quad\quad \,\,\,\,0,\quad\,\,-\ell,\quad -\ell}^{0, \quad -\ell}\right)=|z|^{-\ell}I_\ell(2|z|),\\
 \mathcal{G}_{0}^\ell(1) 
 &=&G_{2,4}^{1,2}
 \left( -|z|^2\big|_{0,\quad \,\,\,\,1,\quad \,\,-\ell,\quad -\ell}^{0, \quad -\ell}\right)=-|z|^{1-\ell}I_{\ell+1}(2|z|),\\
 \mathcal{G}_{0}^\ell(2) 
 &=&G_{2,4}^{1,2}
 \left( -|z|^2\big|_{0,\quad \,\,\,\,2,\quad\quad \,\,-\ell,\quad -\ell}^{0, \quad -\ell}\right)=|z|^{2-\ell}I_{\ell+2}(2|z|).
\end{eqnarray}

\begin{figure}[htbp]
	\resizebox{0.55\textwidth}{!}{
		\includegraphics{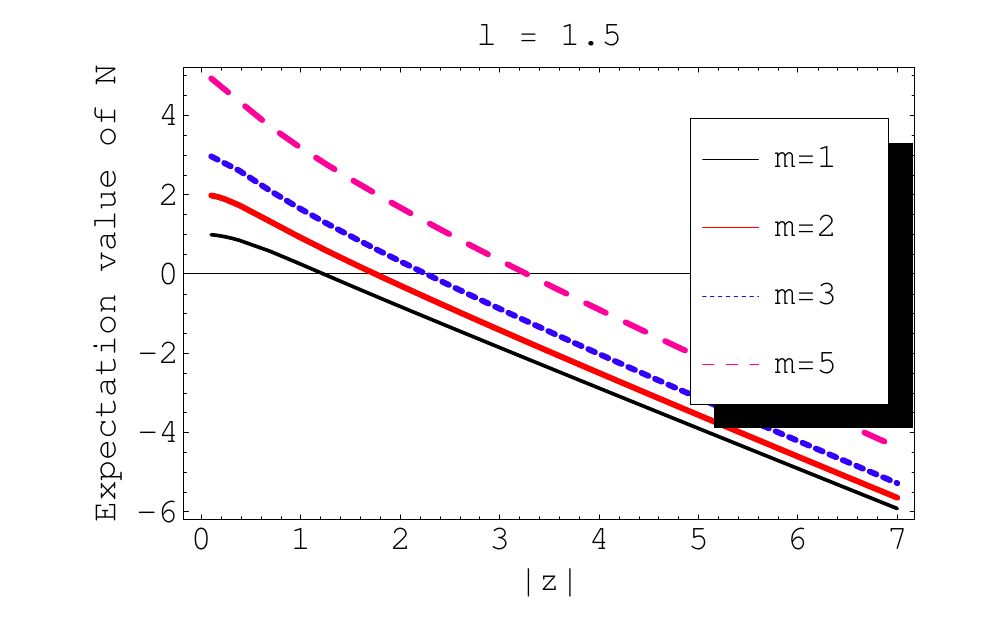}
	}
	\caption{\it \small 
		{  Plots of the expectation values $\langle N\rangle$ (124) versus $|z|$ for different values of
			the photon-number m and with fixed parameter $\ell=1.5$}
	}
	\label{exp}       
\end{figure}

Using the results of the expectation values (\ref{eq5.43}) and (\ref{eq5.431}), we obtain the intensity correlation function and  the Mandel $\mathcal{Q}_{zm}^\ell$-parameter of the PABGCSs as 
\begin{eqnarray} \label{eq5.47}
 g_{\ell, zm}^{(2)}&=& \frac{m(m-1) -2 m \displaystyle{\mathcal{G}_{ m}^\ell(1)\over \mathcal{G}_{ m}^\ell(0)} + \displaystyle{\mathcal{G}_{ m}^\ell(2)\over \mathcal{G}_{ m}^\ell(0)}}{\left(m -\displaystyle {\mathcal{G}_{ m}^\ell(1)\over \mathcal{G}_{ m}^\ell(0)}\right)^2} \\\label{eq5.47a}
 \mathcal{Q}_{zm}^\ell&=& \frac{m^2 -2m +(1 - 2m)\displaystyle{\mathcal{G}_{ m}^\ell(1)\over \mathcal{G}_{ m}^\ell(0)} +
 {\mathcal{G}_{ m}^\ell(2)\over \mathcal{G}_{ m}^\ell(0)}}{m - \displaystyle{\mathcal{G}_{ m}^\ell(1)\over \mathcal{G}_{ m}^\ell(0)}}\label{eq5.48}.
\end{eqnarray} 
For $m=0$, we obtain $g_{\ell, z0}^{(2)}=g_{\ell, z}^{(2)}$ and $\mathcal{Q}_{z0}^\ell=\mathcal{Q}_{z}^\ell$.
In fact the  statistical behaviors of the states $|\psi_{n+m}^\ell\rangle$ are difficult to  guest since the analytical properties of the   intensity correlation function (\ref{eq5.47}) and  the Mandel parameter (\ref{eq5.47a}) depend on the  ratio of Meijer’s $G$-functions. Therefore, we examine the statistical nature  of these states through the numerical computation. Let remark that, the
expectation value of the number operator N (124) vanishes for a certain value
$|z_0|$ depending on the parameters ‘ and m as shown in Figure \ref{exp}.

So the intensity correlation and the Q-Mandel functions are not defined for this value $|z_0|$,
since their denominators depend on this expectation value. We limit then the
analysis of the characteristics of $g_{zm}^{(2)}$  and $\mathcal{Q}_{zm}$ on the domain of sufficently
high values of $|z_0|$ to avoid the undetermined value.

\begin{figure}[htbp]
\resizebox{0.55\textwidth}{!}{
\includegraphics{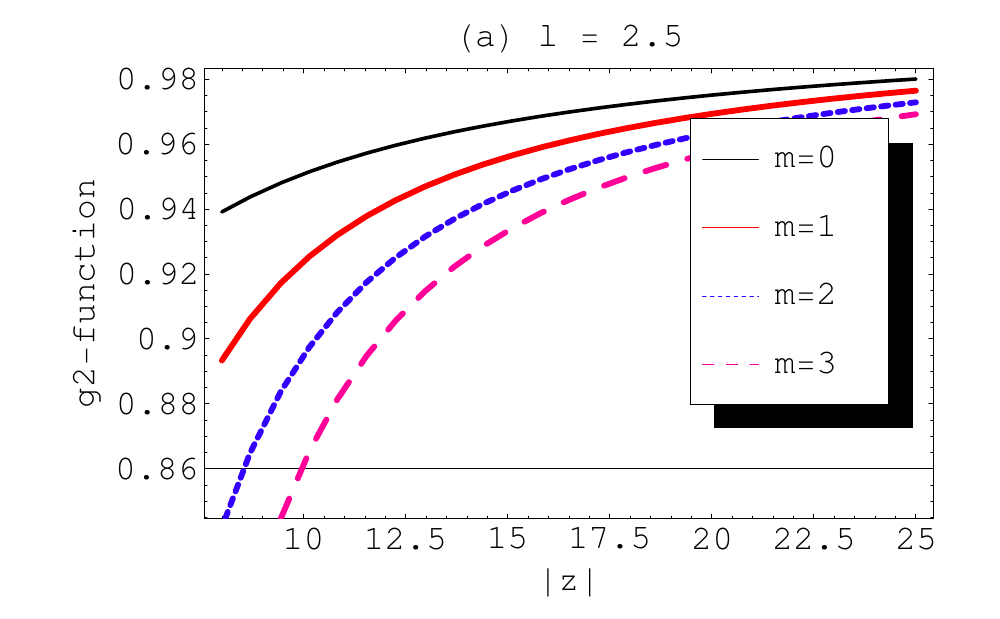}
}
\resizebox{0.55\textwidth}{!}{
\includegraphics{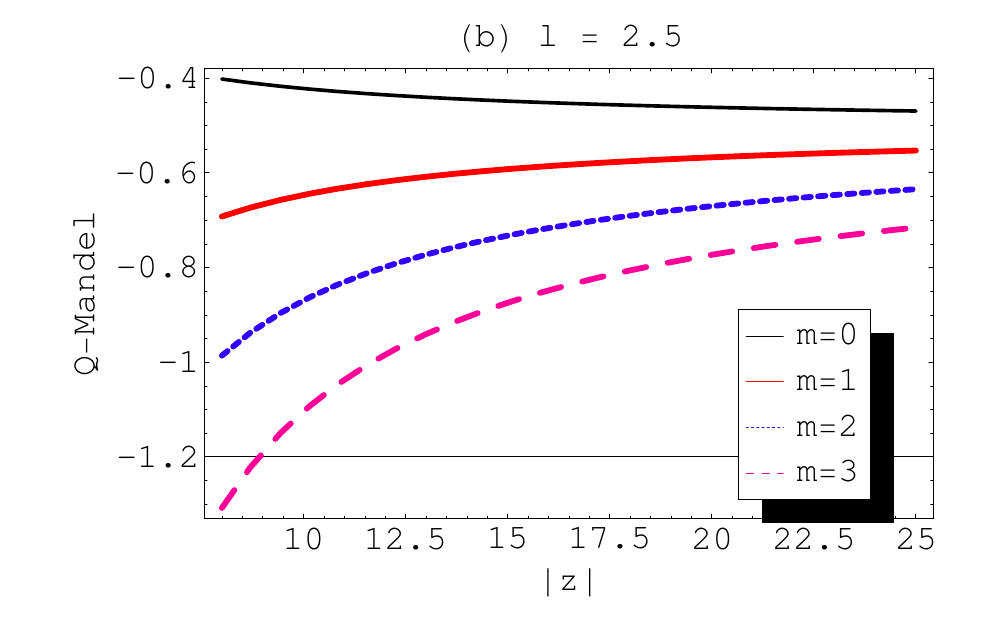}
}
\caption{\it \small  
 Plots of (a)  the intensity corelation function  $g_{\ell, zm}^{(2)}$ (\ref{eq5.47}),  (b) Q-Mandel parameter $\mathcal{Q}_{zm}^\ell$ (\ref{eq5.47a}) of the PABGLCSs versus  $ |z|$ with fixed parameter $\ell = 2.5$ and various values of the  photon-added number $m $.
}
\label{g2-function-PACSTD}       
\end{figure}

 In Figure \ref{g2-function-PACSTD} , we plot (a) the intensity correlation function  (\ref{eq5.47}) and (b) the Q-Mandel parameter  (\ref{eq5.48})  versus   $x = |z|$, with  fixed parameter $ \ell =2.5$  and various values of the  photon-added number $m$. All the curves show that for sufficiently high  values of $|z|$  $g_{\ell, zm}^{(2)}   < 1$ and $  \mathcal{Q}_{zm}^\ell< 0$.  We conclude then that PABGLCSs have also  the sub-Poissonian statistics as the conventional BGLCSs.
   Figures (a) and  (b) show that increasing the photon-added number $m$ decreases both the intensity correlation and the Q-Mandel  functions. So adding more photons increases the depth of the non-classicality of these states. \\\\\

\noindent In a similar manner as  (\ref{eq4.24}), it is possible to determine the  Wigner distribution function corresponding
to the states (\ref{eq5.4}).  Referring to the  Ref.\cite{43}, we obtain:
\begin{eqnarray} \label{eq5.49}
\mathcal{W}_{zm}^\ell(u, p,t)&=&2\frac{[\mathcal{M}_{ m}^\ell (|z|)]^2}{\pi\varpi \hbar}\xi^{\ell - 2m}
\sum_{n'=p'}^\infty \sum_{n=p}^\infty N_{n + m}(\rho,\theta)N_{n' + m}(\rho,\theta)\cr&&\times
e^{i(\gamma_{n + m}^\ell(t)-\gamma_{n' +  m}^\ell(t))}  z^n {z^*}^{n'} \varphi^\ell
\cr&&\times
\sum_{k'=0}^{n'+ m} \sum_{k=0}^{n + m}\binom{n' +  m+\ell}{n' + m-k'}\binom{n + m+\ell}{n + m-k}\cr&& \times\frac{(-\xi)^{k+k'}}{k'!k!}
K_{k-k'-\frac{2ip}{\hbar\varpi}}\left(\varpi \xi\right),
\end{eqnarray}
where 
\begin{eqnarray}
 \varphi^\ell&=&\frac{\Gamma(n'+m+1)\Gamma(n+m+1)}{\Gamma(n'+\ell+1)\Gamma(n+\ell+1)\Gamma(n'+1)\Gamma(n+1)}.
\end{eqnarray}
This result  for the Wigner distribution fits similary with the obtained result of the reference \cite{43}.

\section{Conclusion}\label{section6}
 In the present paper, we have constructed   a new kind  of $SU(1,1)$ coherent states  $\grave{\textrm{a}}$ la Barut-Girardello  deriving from  the  factorisation  method of the TDLP \cite{1}. The minimal set of Klauder's conditions required  to build  coherent states 
 i.e., the label continuity, the normalizability and the overcompleteness \cite{8,9} has been studied and discussed. Statistical properties like  the photon number distribution, the  Mandel parameter and the second-order correlation function are   examined  and analyzed. These  statistical predictions were  confirmed by the  numerical  investigation and we have found that these states are sub-Poissonian in nature. In  addition to these  statistical properties, we have also  examined the behavior of these states  in the phase space from the viewpoint of the Wigner distribution function and  this result is found to  be  similar to the  discussions of the Wigner function of Morse potential \cite{42,43}.
 
  On the other hand,  we have  also  constructed  the $m$-PACSs associated with the BGLCSs of the TDLP. We remarked that both states satisfy also the Klauder's minimum conditions. A comprehensive  study of the statistics of  the $m$-PACSs  are presented  and are expressed  in terms of the modified Bessel function of the third kind, in  Gamma function, and in  Meijer G-function. We have  found   that the addition of photons  from the BGLCSs increases the statistical properties and changes the mathematical properties. 

The  sub-Poissonian statistics of our introduced states are established through  the  nonclassicality sign of the Mandel parameter or the the second-order correlation function.
However, one can also  investigate some others nonclassicality creteria which are usually used in the relevant literatures \cite{44,45}. To achieve this aim, we refer to the normal
squeezing and  amplitude-squared squeezing. In order to study $SU(1,1)$ normal squeezing, we consider the following Hermitian quadrature operators
\begin{eqnarray}
 X_1=\frac{K_++K_-}{2},\,\,\,\,\,\,\, P_1=\frac{K_--K_+}{2i}.
\end{eqnarray}
 The squeezing parameters can be defined as
\begin{eqnarray}
S_\gamma=\frac{\left(\Delta \gamma\right)^2}{\sqrt{\frac{1}{4}|\langle [X_1,P_1]\rangle|^2}}-1,\quad \gamma=X_1,P_1,\quad (\Delta \gamma)^2=\langle \gamma^2\rangle-\langle \gamma\rangle^2.
\end{eqnarray}
 In order, to evaluate  
 $SU(1,1)$ amplitude-squared squeezing, we define the Hermitian operators
 \begin{eqnarray}
 X_2=\frac{K_+^2+K_-^2}{2},\,\,\,\,\,\,\, P_2=\frac{K_-^2-K_+^2}{2i},
\end{eqnarray}
 with the squeezing parameters 
 \begin{eqnarray}
S_l=\frac{\left(\Delta l\right)^2}{\sqrt{\frac{1}{4}|\langle [X_2,P_2]\rangle|^2}}-1,\quad l=X_2,P_2, \,\mbox{with}\quad (\Delta l)^2=\langle l^2\rangle-\langle l\rangle^2.
\end{eqnarray}
The continuation of these computations are under investigation, we hope to report these aspects elsewhere.

	\section*{Acknowledgments}
	L.M Lawson acknowledges support from AIMS-Ghana
	under the Postdoctoral fellow/teaching assistance (Tutor) grant

\end{document}